# Coherent vector π-pulse in optical amplifiers

**Elena V. Kazantseva[a], Andrei I. Maimistov[b], Sergei O. Elyutin[b], and Stefan Wabnitz[a]**

[a]Laboratoire de Physique de l'Université de Bourgogne, UMR CNRS 5027, 9 Avenue

Alain Savary, BP 47 870,  21078 Dijon, France;

..[b]Moscow Engineering Physics Institute, Kashirskoe sh. 31, Moscow, 115409 Russia

Abstract

We obtain an exact vector solitary solution for the amplification of an optical pulse with a time width short compared with both population and polarization decay time. This dissipative soliton results from the balance between the gain from inverted resonant two-level atoms and the linear loss of the host material.  We suppose that the excited state of the active centers is degenerate on the projection of the angular moment. Numerical simulations demonstrate the stability of these vector dissipative solitons in the presence of both linear birefringence and group velocity dispersion of the host material.

## Introduction

The first and perhaps most striking manifestation of the soliton concept in optics is the self-induced transparency effect discovered in 1967 by Mc Call and Hahn, whereby a short optical pulse may penetrate undistorted through an absorbing medium [1]. The mechanism for such a fascinating phenomenon is the coherent equilibrium between absorption at the leading edge and stimulated emission at the trailing edge of a so-called 2-$\pi$ pulse, which leads to the slowing down of the pulse. A perhaps less known but equally interesting effect was pointed out in 1965 by Arecchi and Bonifacio [2], who discussed the creation of dissipative solitons in optical laser amplifiers. This solution is known as $\pi$-pulse, and it represents the coherent balance between the stimulated emission of an inverted atomic population and the linear loss of the host material [3]. Experiments of coherent pulse propagation in bulk media are hampered by the spatial spreading of the beams due to diffraction [1,4-5].

This limitation can be lifted when guiding the light pulses in an optical fiber containing resonant impurities. Indeed, the observation of self-induced transparency and $\pi$-pulse soliton propagation over several meters was carried out by Nakazawa et al. by means of an erbium-doped optical fiber [6-7]. In these experiments, it was necessary to cool down the fiber to a few Kelvin in order to increase the homogeneous polarization lifetime $T_2$ up to about 10 ns. Indeed, at room temperature the transition of erbium which is resonant with 1500 nm light has a lifetime $T_2$ of 250 fs [8]. For femtosecond pulse durations, the non-resonant, instantaneous Kerr nonlinear response of the host silica fiber is more than six orders of magnitude larger than the resonant response [6-7], which hinders in practice coherent pulse propagation effects [9-13].

The recent technological development of photonic bandgap (PBG) optical fibers, where light is guided inside an hollow core by the mechanism of Bragg reflection, permits to reduce the Kerr nonlinearity of the host material by several orders of magnitude with respect to conventional index guiding silica core fibers. Filling the core of the PBG fiber with a resonant atomic or molecular gas, makes it practical to observe coherent wave propagation effects such as electromagnetically-induced transparency over relatively long lengths [14-15].

In this work, we analyze the coherent propagation of short polarized pulses in a lossy host fiber filled with active resonant two-level atoms. We neglect for simplicity the presence of the Kerr nonlinear response of the host medium, which is a legitimate hypothesis in PBG fibers. On the other hand, we include in our treatment the linear loss, birefringence and group-velocity dispersion (GVD) of the host fiber. In the case of a short optical pulse that propagates at the zero GVD wavelengths in an isotropic, lossy PBG fiber with resonant active impurities, we have been able to obtain a simple vector $\pi$-pulse soliton solution to the coherent pulse amplification problem.

The paper has the following structure. Section 2 is devoted to the formulation of the problem. The analytical solution to the propagation of a polarized pulse is obtained in section 3. In section 4 we numerically study the robustness of this vector $\pi$-pulse propagating in the amplifier in the presence of host fiber linear birefringence and group velocity dispersion.

## Basic equations

Let us consider the propagation of ultra-short optical pulse envelopes of arbitrary polarization in a birefringent host optical fiber, doped by two-level atoms whose energy transition is resonant with the optical carrier frequency. Pulse propagation in the fiber may be described in terms of the reduced Maxwell equations, coupled with the generalized Bloch equations [16-20] that determine the evolution of the doping atoms.

The polarized electric field of a short optical pulse that propagates along the z axis of the fiber may be written as $\mathbf{E} = f_x(\rho)E_x\mathbf{e}_x + f_y(\rho)E_y\mathbf{e}_y$, where $\mathbf{e}_x$ and $\mathbf{e}_y$ are orthogonal unit vectors in the $x$ and $y$ directions. Moreover, $f_{x,y}(\rho)$ represents the transverse profile of the two orthogonally polarized fundamental modes of the fiber. In the frame of the usual slowly varying (complex) envelope approximation (SVEA), we may write

$$E_x = \mathsf{E}_x(z,t)\exp[i(\beta_x z - \omega_0 t)], \;\; E_y = \mathsf{E}_y(z,t)\exp[i(\beta_y z - \omega_0 t)],$$

where $\omega_0$ is the carrier angular frequency, $\beta_x(\beta_y)$ is the linear propagation constant of the slow (fast) mode of the birefringent fiber.

We may write the propagation constants of the two orthogonal fiber modes in terms of their average value $\beta$ and birefringence $\Delta\beta$, so that $\beta_x = \beta + \Delta\beta$ and $\beta_y = \beta - \Delta\beta$. The slowly varying complex envelopes of the field may be expressed in terms of real amplitudes $R_{x,y}$ and phases $\phi_{x,y}$, that is $\mathsf{E}_x = R_x \exp[i\varphi_x]$, $\mathsf{E}_y = R_y \exp[i\varphi_y]$. Let us express the phases as $\varphi_x = \tilde{\varphi} + \phi$ and $\varphi_y = \tilde{\varphi} - \phi$ (where $\tilde{\varphi}$ is an average phase), so that the two components of the electric field read as

$$E_x = A_x(z,t)\exp[i\beta z - i\omega_0 t], \;\; E_y = A_y(z,t)\exp[i\beta z - i\omega_0 t],$$

where

$$A_x = R_x \exp[i(\tilde{\varphi} + \phi + \Delta\beta z)], \text{ and } A_y = R_y \exp[i(\tilde{\varphi} - \phi - \Delta\beta z)].$$

In the above expression, the linear birefringence of the fiber is represented by the phase terms $\pm\Delta\beta z$. Whereas the presence of the doping atoms leads to a nonlinear birefringence which is represented by the terms $\pm\phi(t,z)$. In this paper we shall ignore the presence of the intrinsic nonlinear response (or third order susceptibility) of the host fiber that contains the resonant impurities. This hypothesis can be justified *a posteriori* by verifying that the nonlinear length associated with the intensity dependent refractive index of the host fiber is much longer than the distance for the manifestation of the coherent propagation effects that are of interest here. Indeed, the nonlinear index coefficient of photonic bandgap optical fibers is several orders of magnitude smaller than for standard silica fibers.

In the frame of the SVEA, one obtains from the Maxwell equations (with a source term that accounts for the resonant polarization from the doping atoms) the following coupled equations for the amplitudes $A_x$ and $A_y$

$$i\frac{\partial A_x}{\partial z} + iv_x^{-1}\frac{\partial A_x}{\partial t} - \sigma_x\frac{\partial^2 A_x}{\partial t^2} + \Delta\beta A_x + \kappa P_x + i\gamma A_x = 0 \ , \qquad (1.1)$$

$$i\frac{\partial A_y}{\partial z} + iv_y^{-1}\frac{\partial A_y}{\partial t} - \sigma_y\frac{\partial^2 A_y}{\partial t^2} - \Delta\beta A_y + \kappa P_y + i\gamma A_y = 0 \ . \qquad (1.2)$$

In the above equations (1), the action of the resonant impurities is described by the slowly varying polarization envelopes $P_x$ and $P_y$. Moreover, in Eqs.(1) the coefficient $\gamma$ describes linear fiber loss, whereas the group velocities and dispersions of the two orthogonal fiber modes are $v_{x,y}^{-1} = d\beta_{x,y}/d\omega$ and $\sigma_{x,y} = (1/2)d^2\beta_{x,y}/d\omega^2$, respectively. In addition, the coupling coefficient $\kappa = 2\pi\omega_0 n_a / cn(\omega_0)$, where $n_a$ is the concentration of the impurity atoms.

As we shall see, Eqs. (1) possess a solitary wave solution as result of the balance between linear fiber loss and gain from the doping resonant impurities with pumping. In Eqs. (1), the electric field of the pulse is expressed in terms of Cartesian components. When considering the presence of the nonlinear polarization, it proves convenient to express Eqs. (1) in terms of right and left-hand circularly polarized $E_{1,2}$ components, with $E_1 = E_x + iE_y$ and $E_2 = E_x - iE_y$. The corresponding complex envelopes of the electric field and the polarization may be written as $A_1 = A_x + iA_y$, $A_2 = A_x - iA_y$, $P_1 = P_x + iP_y$, and $P_2 = P_x - iP_y$. Indeed, as discussed below the eigenmodes of the nonlinear birefringence exhibit a circular state of polarization.

Let us complement Eqs. (1) for the optical field with the equations that describe the temporal dynamics of the density matrix of the two-level atom ensemble embedded in the fiber host material. We shall suppose that the energy levels of these atoms are degenerate over the projections of the angular momentum [21-23], $j_a$ and $j_b$, where the subscript $a$ and $b$ denote the upper and the lower energy states, respectively. For the sake of definiteness, we shall consider in this work the case where $j_a = 1 \rightarrow j_b = 0$. The resonant atomic transition is characterised by the dipole momentum operator elements $d_{13} = d_{23} = d_{31}^* = d_{32}^* = d$. Let us derive now the system of generalized Bloch equations that describe the time evolution of the resonant impurities. We introduce the following notation for the slowly varying elements of the density matrix $\hat{\rho}$ that describe the transition between the two states $|a, m> = |j_a = 1, m = \pm 1>$ and $|b> = |j_b = 0, m = 0>$

$$\rho_{12} = <a, -1|\hat{\rho}|a, +1>, \quad \rho_{13} = <a, -1|\hat{\rho}|b>, \quad \rho_{23} = <a, +1|\hat{\rho}|b>,$$
$$\rho_{11} = <a, -1|\hat{\rho}|a, -1>, \quad \rho_{22} = <a, +1|\hat{\rho}|a, +1>, \quad \rho_{33} = <b|\hat{\rho}|b>,$$
$$\rho_{kl} = \rho_{lk}^*, \quad l, k = 1, 2, 3$$

We shall restrict our attention in this work to case of optical pulse durations which are much shorter than all relaxation times involved in the resonant atomic subsystem. Under this assumption, the system of generalized Bloch equations may be written in terms of the right-hand and left-hand circularly polarized field components as follows

$$i\hbar \frac{\partial \rho_{13}}{\partial t} = \hbar \Delta \omega \rho_{13} - d_{13}(\rho_{33} - \rho_{11})A_1 + d_{23}\rho_{12}A_2, \quad (2.1)$$

$$i\hbar\frac{\partial\rho_{23}}{\partial t} = \hbar\Delta\omega\rho_{23} - d_{23}(\rho_{33} - \rho_{22})A_2 + d_{13}\rho_{21}A_1, \qquad (2.2)$$

$$i\hbar\frac{\partial\rho_{12}}{\partial t} = -d_{13}\rho_{32}A_1 + d_{32}\rho_{13}A_2^*, \qquad (2.3)$$

$$i\hbar\frac{\partial}{\partial t}(\rho_{33} - \rho_{11}) = 2\left(d_{13}\rho_{31}A_1 - d_{31}\rho_{13}A_1^*\right) + \left(d_{23}\rho_{32}A_2 - d_{32}\rho_{23}A_2^*\right), \qquad (2.4)$$

$$i\hbar\frac{\partial}{\partial t}(\rho_{33} - \rho_{22}) = \left(d_{13}\rho_{31}A_1 - d_{31}\rho_{13}A_1^*\right) + 2\left(d_{23}\rho_{32}A_2 - d_{32}\rho_{23}A_2^*\right). \qquad (2.5)$$

The boundary (i.e., at $t \to -\infty$) conditions for the non-diagonal elements of the density matrix $\hat{\rho}$ are $\rho_{12} = \rho_{13} = \rho_{23} = 0$. Whereas two possibilities exist for the boundary conditions associated with the diagonal matrix elements of the density matrix $\hat{\rho}$, depending on the polarization state of the pump. The case of a linearly polarized pump is characterized by the boundary conditions $\rho_{33} = 0, \rho_{22} = \rho_{11} = 1/2$. Whereas with a circularly polarized pump one obtains $\rho_{33} = 0, \rho_{22} = 0, \rho_{11} = 1$ or $\rho_{33} = 0, \rho_{22} = 1, \rho_{11} = 0$, depending on the handedness of the pump wave.

The relationship between the polarization source components and the elements of the density matrix that describe the evolution of the resonant impurities is given by the following expressions

$$\kappa P_1 = \frac{2\pi\omega_0 n_{at} d_{13}}{cn(\omega_0)}\langle\rho_{13}\rangle, \quad \kappa P_2 = \frac{2\pi\omega_0 n_{at} d_{13}}{cn(\omega_0)}\langle\rho_{23}\rangle.$$

where $\langle\ \rangle$ indicates summation over all atoms within a frequency detuning $\Delta\omega t_0$, measured from the center of the atomic transition with inhomogeneous broadening [17]. Hereafter we shall assume for simplicity that the inhomogeneous broadening is absent, that is we will work under the hypothesis of a sharp atomic resonant transition.

In order to simplify our notation, which is important when performing numerical simulations, we shall adopt dimensionless units. Namely, we set

$$A_{1,2} = A_0 e_{1,2} , \; z = \zeta L , \; \tau = (t - z / v) t_0^{-1} ,$$

where $t_0$ an arbitrary characteristic time scale (it can be set for example equal the initial pulse duration $t_{p0}$). Moreover, $L = \gamma^{-1}$ and $A_0 = \hbar (t_0 \mid d_{13} \mid)^{-1}$ denote the linear absorption length and a characteristic amplitude, and we adopted a time retarded frame, moving at the mean velocity $v$ of the two polarization components of an optical pulse, i.e., $v^{-1} = (v_x^{-1} + v_y^{-1})/2$. The system of equations (1) and (2) in terms of new variables reads:

$$i \frac{\partial e_1}{\partial \zeta} + \frac{i}{l_g} \frac{\partial e_2}{\partial \tau} - \frac{s}{l_d} \frac{\partial^2 e_1}{\partial \tau^2} + \frac{1}{l_c} e_2 + \frac{1}{l_a} \langle \rho_{13} \rangle + i e_1 = 0 , \qquad (3.1)$$

$$i \frac{\partial e_2}{\partial \zeta} - \frac{i}{l_g} \frac{\partial e_1}{\partial \tau} - \frac{s}{l_d} \frac{\partial^2 e_2}{\partial \tau^2} + \frac{1}{l_c} e_1 + \frac{1}{l_a} \langle \rho_{23} \rangle + i e_2 = 0 . \qquad (3.2)$$

$$i \frac{\partial \rho_{13}}{\partial \tau} = \Delta \rho_{13} + (\rho_{11} - \rho_{33}) e_1 + \rho_{12} e_2 , \qquad (4.1)$$

$$i \frac{\partial \rho_{23}}{\partial \tau} = \Delta \rho_{23} + (\rho_{22} - \rho_{33}) e_2 + \rho_{12}^* e_1 , \qquad (4.2)$$

$$i \frac{\partial \rho_{12}}{\partial \tau} = -\rho_{23}^* e_1 + \rho_{13} e_2^* , \qquad (4.3)$$

$$\frac{\partial}{\partial \tau} (\rho_{33} - \rho_{11}) = -4 \operatorname{Im}(\rho_{13} e_1^*) - 2 \operatorname{Im}(\rho_{23} e_2^*) , \qquad (4.4)$$

$$\frac{\partial}{\partial \tau} (\rho_{33} - \rho_{22}) = -2 \operatorname{Im}(\rho_{13} e_1^*) - 4 \operatorname{Im}(\rho_{23} e_2^*) , \qquad (4.5)$$

where $\Delta = \Delta \omega t_0$. In Eqs.(4), the dimensionless distances $l_g, l_d, l_c,$ and $l_a$ are defined as

$$l_g^{-1} = \left(v_x^{-1} - v_y^{-1}\right)/2\gamma_0 \,, \ l_d^{-1} = |\sigma|/\gamma_0^2 \,, \ l_c^{-1} = \Delta\beta/\gamma \,, \ l_a^{-1} = (2\pi\omega_0 t_0 n_{at} \mid d_{13}\mid^2)/(\gamma cn(\omega_0)\hbar) \,,$$

where we took for simplicity equal dispersion $\sigma_x = \sigma_y = \sigma$ on both fiber axes, s=sgn($\sigma$), and we suppose $v_y > v_x$. The resonant absorption length $l_a$ measures the ratio between coherent and incoherent absorption lengths in the doped fiber. Whereas the quantity $l_c$ represents the ratio between the birefringence-induced rotation distance of circular polarizations and the linear absorption length $L = \gamma^{-1}$. Indeed, in Eqs. (3) birefringence introduces the linear coupling between right and left circular components of the propagating pulse. Finally, the difference between the group velocities of the two linearly polarized modes of the host fiber leads to polarization mode dispersion (of temporal walk-off) over the characteristic distance $l_g$. By neglecting the frequency dependence of the index difference between the x and y polarized modes, one obtains $l_c / l_g \approx \lambda / 2\pi(ct_{p0})$. Therefore, for pulse widths in the picosecond range it is possible to neglect the pulse walk-off as it occurs over a distance which is much longer than the linear beat-length of the fiber. On the other hand, for femtosecond pulses it is necessary to take into account the effects of polarization mode dispersion. For example, for a signal pulse at the carrier wavelength $\lambda = 1500 nm$ and time duration $t_{p0} = 10$ fs, the ratio $l_c / l_g \approx 0.08$.

The vector Maxwell-Bloch equations (3) and (4) provide the basis for describing the propagation of short polarized pulses of light in a lossy, birefringent fiber, doped by active resonant impurities.

## Exact vector $\pi$-pulse

It has been known for a long time that the balance between linear absorption and nonlinear gain resulting from an inverted two-level atomic medium leads to a stable steady-state pulse in a laser amplifier. Such a pulse was named the $\pi$-pulse [2-3]. Therefore, we shall name as "vector $\pi$-pulse" the solitary wave which results as a balance between the linear absorption of the host material and the coherent amplification of the doping atoms. In order to find a simple analytic expression for the vector $\pi$-pulse, we shall neglect at first the presence of group velocity dispersion in the fiber, that is we set $l_d \gg L,\ l_a$.

Let us consider at first the case of a nearly isotropic fiber, that is a situation where the linear birefringence of the fiber can be neglected over both distances $L$. In other words, we assume a relatively strong interaction of the optical pulse with the resonant atomic medium, so that the coupling length $l_c \gg l_a$. In this case, we may reduce Eqs. (3) to the following equations

$$i\frac{\partial e_1}{\partial \zeta} + \frac{1}{l_a}\rho_{13} + ie_1 = 0 \qquad (5.1)$$

$$i\frac{\partial e_2}{\partial \zeta} + \frac{1}{l_a}\rho_{23} + ie_2 = 0, \qquad (5.2)$$

Quite remarkably, Eqs.(4-5) admit of a simple analytical solution. Following refs. [2-3], a scalar $\pi$-pulse in a two-level laser amplifier with linear dissipation is a solitary wave that propagates with the speed of light in the medium, i.e., $v$ in the present notation. Therefore, in the search for a steady-state solution of Eqs.(4-5) that represents vector $\pi$-pulses (pulse of the polarized radiation) one may set $\partial e_{1,2}/\partial \zeta = 0$ and $\Delta\omega = 0$ in the system (4). In this case, from Eqs.(5) it follows that

$$e_1 = il_a^{-1}\rho_{13},\ \ e_2 = il_a^{-1}\rho_{23}, \qquad (6)$$

and Bloch equations (4) reduce to

$$i\frac{\partial \rho_{13}}{\partial \tau} = -n_1 e_1 + \rho_{12} e_2, \qquad i\frac{\partial \rho_{23}}{\partial \tau} = -n_2 e_2 + \rho_{12}^* e_1, \qquad (7.1)$$

$$i\frac{\partial \rho_{12}}{\partial \tau} = -\rho_{23}^* e_1 + \rho_{13} e_2^*, \qquad (7.2)$$

$$\frac{\partial}{\partial \tau} n_1 = -4\,\text{Im}\left(\rho_{13} e_1^*\right) - 2\,\text{Im}\left(\rho_{23} e_2^*\right), \qquad \frac{\partial}{\partial \tau} n_2 = -2\,\text{Im}\left(\rho_{13} e_1^*\right) - 4\,\text{Im}\left(\rho_{23} e_2^*\right) \qquad (7.3)$$

where $n_1 = \rho_{33} - \rho_{11}$, $n_2 = \rho_{33} - \rho_{22}$. Let us define $e_1 = a_1 \exp(i\varphi_1)$ and $e_2 = a_2 \exp(i\varphi_2)$, and we set $\rho_{13} = -iu \exp(i\varphi_1)$ and $\rho_{23} = -ir \exp(i\varphi_2)$. The right-hand side of equation (7.2) reads as

$$-\rho_{23}^* e_1 + \rho_{13} e_2^* = -i(ra_1 + ua_2)\exp\{i(\varphi_1 - \varphi_2)\}.$$

Therefore we may set $\rho_{12} = (p + is)\exp\{i(\varphi_1 - \varphi_2)\}$. By using these definitions, we may represent the system of equations (7) in a new form involving the real variables $u$, $r$, $n_1$, $n_2$, $p$, $s$, $\phi_1$, and $\phi_2$

$$\frac{\partial u}{\partial \tau} = -a_1 n_1 + pa_2, \qquad \frac{\partial r}{\partial \tau} = -a_2 n_2 + pa_1, \qquad (8.1)$$

$$\frac{\partial n_1}{\partial \tau} = 4a_1 u + 2a_2 r, \qquad \frac{\partial n_2}{\partial \tau} = 2a_1 u + 4a_2 r, \qquad (8.2)$$

$$\frac{\partial p}{\partial \tau} = s\frac{\partial}{\partial \tau}(\varphi_1 - \varphi_2) - (ra_1 + ua_2), \qquad (8.3)$$

$$u\frac{\partial \varphi_1}{\partial \tau} - sa_2 = 0, \qquad r\frac{\partial \varphi_2}{\partial \tau} + sa_1 = 0 \qquad (8.4)$$

$$\frac{\partial s}{\partial \tau} = p\frac{\partial}{\partial \tau}(\varphi_1 - \varphi_2), \qquad (8.5)$$

Whenever $s = 0$ and $p = 0$ for $\tau \to -\infty$, from (8.5) it follows that $\partial s / \partial \tau = 0$ for $\tau \to -\infty$, therefore $s \equiv 0$. Hence, according to (8.4), the time derivatives of the optical pulse phases are also identically equal to zero, that is $\partial \varphi_1 / \partial \tau \equiv 0$ and $\partial \varphi_2 / \partial \tau \equiv 0$. The system of equations (8) reduces to

$$\frac{\partial u}{\partial \tau} = -a_1 n_1 + pa_2, \qquad \frac{\partial r}{\partial \tau} = -a_2 n_2 + pa_1 \qquad (9.1)$$

$$\frac{\partial n_1}{\partial \tau} = 4a_1 u + 2a_2 r, \quad \frac{\partial n_2}{\partial \tau} = 2a_1 u + 4a_2 r \qquad (9.2)$$

$$\frac{\partial p}{\partial \tau} = -(ra_1 + ua_2). \qquad (9.3)$$

With the help of the substitution $a_1 = l_a^{-1} u$, $a_2 = l_a^{-1} r$, one obtains from Eqs.(9)

$$l_a \frac{\partial u}{\partial \tau} = -un_1 + pr, \quad l_a \frac{\partial r}{\partial \tau} = -rn_2 + pu \qquad (10.1)$$

$$l_a \frac{\partial n_1}{\partial \tau} = 4u^2 + 2r^2, \quad l_a \frac{\partial n_2}{\partial \tau} = 2u^2 + 4r^2 \qquad (10.2)$$

$$l_a \frac{\partial p}{\partial \tau} = -2ur. \qquad (10.3)$$

Let us consider now the particular case of a pump pulse with linear polarization. In this case, we have $n_1 = n_2 = n$, $r = u$, and the initial (i.e., for $\tau \to -\infty$) population of the resonant atoms is equal to $n = n_0 = -1/2$. From Eqs.(10), one obtains

$$l_a \frac{\partial u}{\partial \tau} = -un + pr, \qquad (11.1)$$

$$l_a \frac{\partial p}{\partial \tau} = -2u^2, \qquad (11.2)$$

$$l_a \frac{\partial n}{\partial \tau} = 6u^2. \qquad (11.3)$$

From Eqs.(11.2) and (11.3), one obtains the first integral

$$3p + n = n_0. \qquad (12.1)$$

Additionally, from Eq. (11) one finds the second integral of motion

$$6u^2 + n^2 + 3p^2 = n_0^2. \qquad (12.2)$$

By using the results of Eqs.(12), equation (11.3) can be rewritten as

$$l_a \frac{\partial n}{\partial \tau} = n_0^2 - n^2 - \frac{1}{3}(n_0 - n)^2 \ ,$$

and, since $n_0 = -1/2$, one obtains

$$l_a \frac{\partial n}{\partial \tau} = \frac{1}{4} - n^2 - \frac{1}{3}\left(\frac{1}{2} + n\right)^2 = -\frac{1}{3}\left(4n^2 + n - \frac{1}{2}\right), \tag{13}$$

The above equation can be also expressed in the form

$$l_a \frac{\partial n}{\partial \tau} = -\frac{1}{3}\left[\left(2n + \frac{1}{4}\right)^2 - \frac{9}{16}\right]. \tag{14}$$

Let us introduce now the new variable

$$w = (4/3)(2n + 1/4) \ ,$$

so that $\partial w/\partial \tau = l_a^{-1}(1 - w^2)/2$. The solution of the differential equation for $w$ is $w = \tanh \theta$, with $\theta = l_a^{-1}(\tau - \tau_0)/2$. Hence, the solution of Eq.(13) can be written as

$$n(\tau) = (3 \tanh \theta - 1)/8 \ . \tag{15.1}$$

From the above expression, one may verify that, whenever $\tau \to -\infty$, $n \to -1/2$. Nevertheless, whenever $\tau \to +\infty$, one obtains that $n \to 1/4$. In other words, contrary to the case of a scalar $\pi$ pulse [2-3], in the present case the atomic system does not reach the ground state after the interaction with the polarized optical pulse.

From Eqs.(15.1) and (11.3), we obtain the envelope of the dissipative soliton pulse as

$$u = (4\sqrt{2})^{-1} \operatorname{sech}[l_a^{-1}(\tau - \tau_0)/2].$$

Moreover,

$$p(\tau) = -\frac{1}{8}\left[1 + \tanh\left(l_a^{-1}(\tau - \tau_0)/2\right)\right], \tag{15.2}$$

$$a_1(\tau) = a_2(\tau) = (4\sqrt{2}l_a)^{-1}\,\text{sech}[l_a^{-1}(\tau - \tau_0)/2]. \qquad (15.3)$$

It is worth noting from Eq.(15.2) that, whenever $\tau \rightarrow +\infty$, one obtains $\text{Re}\,\rho_{23} = p \rightarrow -1/4$, that is the matrix element $\rho_{23}$ does not return to zero. In Figure 1 we display the amplitude of the two polarization components of the linearly polarized vector $\pi$-pulse over the total propagation distance $\zeta$ =25.

In the case of a circularly polarized pump beam, one obtains the following boundary conditions: $\rho_{33} = 0, \rho_{22} = 0, \rho_{11} = 1$, or $\rho_{33} = 0, \rho_{22} = 1, \rho_{11} = 0$, depending on the pump handedness. In this case, the system of equations (7) reduces to the usual Bloch equations for a two-level system. In this case, the solution of Eqs.(6-7) with the boundary conditions $\rho_{33} = 0, \rho_{22} = 0, \rho_{11} = 1$, leads to the soliton solution

$$u = 1/\cosh[l_a^{-1}(\tau - \tau_0)], \; n_1 = \tanh[l_a^{-1}(\tau - \tau_0)] \; \; p(\tau) = 0$$

$$a_1 = l_a^{-1}/2\cosh[l_a^{-1}(\tau - \tau_0)], \; a_2 = 0 \,.$$

The above solution corresponds to a right-handed circularly polarized $\pi$-pulse, which is fully equivalent to the well-known scalar $\pi$-pulse solution of a coherent laser amplifier. Clearly, with the boundary conditions $\rho_{33} = 0, \rho_{22} = 1, \rho_{11} = 0$, the solution of Eqs. (6-7) is a left-handed circularly polarized $\pi$-pulse.

## Robustness of vector $\pi$-pulses

In the previous section we demonstrated that, in the case of an isotropic, non dispersive optical fiber, a linearly polarized optical pulse may propagate undistorted in a two-level resonant coherent amplifier as a result of a balance between host fiber loss and gain from the doping atoms. However, in real fibers there is always some degree of linear birefringence $\Delta\beta$ between the two orthogonal linear polarizations. Moreover, unless the carrier frequency of the vector $\pi$-pulse is placed exactly at the zero dispersion value of the fiber, GVD may play a significant role in the temporal evolution of a short optical pulse.

In this section we shall examine, by means of numerical simulations, the robustness of vector $\pi$-pulse creation and propagation in the presence of either linear birefringence or GVD of the host fiber. The electric field equations (3) were solved by means of the finite-difference implicit-explicit Crank–Nicolson numerical scheme [24], and the predictor–corrector procedure was applied to compute the numerical solution of the Bloch equations (4). In a fiber with linear birefringence (and no GVD), the reduced Maxwell equations (3) read as

$$i\frac{\partial e_1}{\partial \zeta} + \frac{1}{l_c}e_2 + \frac{1}{l_a}\rho_{13} + ie_1 = 0, \tag{16.1}$$

$$i\frac{\partial e_2}{\partial \zeta} + \frac{1}{l_c}e_1 + \frac{1}{l_a}\rho_{23} + ie_2 = 0, \tag{16.2}$$

whereas the resonant atomic system evolves according to the Bloch equations (7). Let us consider the influence of linear birefringence on the stability of the vector $\pi$-pulse solutions (15). In order to vary the angle with respect to the fiber axes of the input linearly polarized pulse, we introduced a constant phase shift $\vartheta$ between the circular components, i.e., $e_1(\zeta = 0, \tau) = a_1(\tau)$, $e_2(\zeta = 0, \tau) = \exp(i\vartheta)a_1(\tau)$. We performed numerical simulations of the amplification and propagation of these initial pulses for different values of the phase $\vartheta$. Unless $\vartheta$ is equal to $\pi n$, where $m = 0,1,2,\ldots$ (i.e., the input pulse is aligned with either the slow or the fast axis of the fiber, respectively), linear birefringence leads to the spatially periodic (with period equal to the beat length $\pi l_c$) rotation on the Poincaré sphere of the polarization vector of the $\pi$-pulse about the principal axes of the fiber. This rotation involves the periodic energy exchange between the circular components of the vector π-pulse which is shown in figure 2 for $\vartheta = \pi/2$. In these simulations we set the birefringence length $l_c = 2$, whereas the resonant absorption distance is $l_a = 0.5$. Figure 3 shows that by decreasing the birefringence length down to $l_c = 0.5$, one still observes a stable rotation of the polarization state in spite of the relatively short value of the linear beat length.

Let us consider next the stability of the propagation of the vector π-pulse (15) in the presence of host group velocity dispersion. To this end, we neglected for simplicity the presence of fibre birefringence and numerically solved the coupled equations

$$i\frac{\partial e_1}{\partial \zeta} - \frac{s}{l_d}\frac{\partial^2 e_1}{\partial \tau^2} + \frac{1}{l_a}\rho_{13} + ie_1 = 0 \ , \tag{17.1}$$

$$i\frac{\partial e_2}{\partial \zeta} - \frac{s}{l_d}\frac{\partial^2 e_2}{\partial \tau^2} + \frac{1}{l_a}\rho_{23} + ie_2 = 0 \ . \tag{17.2}$$

along with the Bloch equations (7), with $l_a = 0.5$, $s = -1$ (anomalous dispersion regime), and for different values of the dispersion distance $l_d$. In Figure 6 we show that for weak host GVD (i.e., $l_d = 5$) the propagation of the vector π-pulse is virtually unaffected over a distance up to several tens of the linear absorption length. On the other hand, quite remarkably Figure 7 shows that, even in the presence of relatively strong host GVD (here $l_d = 2$ $l_a = 1$), light in the vector π-pulse remains temporally confined in spite of its substantial reshaping at $\zeta = 10$. A typical estimation of the soliton pulse amplitude is $A \approx 0.16 \cdot 10^7 \, V/m$ for pulse widths $\approx 6.77 \cdot ps$. This amplitude value should be compared with the atomic electric field, which is equal to $A_{at} \approx 5 \cdot 10^{11} \, V/m$.

## Excitation of vector p-pulse

In the previous sections we studied the propagation stability of vector π-pulses in the presence of perturbations such as host birefringence or GVD, with an input optical pulse given by the solution (15). For the practical observation of vector π-pulse solitons, it is important to consider the stability of the generation of these pulses whenever the input optical pulse does not coincide with the exact soliton solution. In fact, a generic property of stable dissipative solitons is their capability to attract pulses of initial different amplitudes and time widths. To this end, we investigated by extensive numerical simulations the temporal dynamics of the formation of steady-state pulses in the coherent resonant amplifier with linear losses. Let us consider a linearly polarized pumping of the doping atoms, i.e., we set $\rho_{33} = 0$, $\rho_{22} = \rho_{11} = 1/2$. Moreover, we set the input optical pulse as

$$e_1(\zeta = 0, \tau) = e_{01}\mathrm{sech}(\tau/\tau_0), \quad e_2(\zeta = 0, \tau) = e_{02}\mathrm{sech}(\tau/\tau_0). \qquad (18)$$

A particularly interesting case is represented by the possibility of generating a vector π-pulse by the coherent amplification of a relatively small amplitude input pulse. Numerical simulations of Eqs. (7) and (16) show that, in the absence of linear birefringence, equal weak initial optical pulses transform into equal amplitude steady-state solitons, with asymptotic level populations $n_{11,}n_{22} \to 1/4$ at $\tau \to +\infty$. In the case of unequal amplitudes of the two circular polarization components of the initial weak pulse, numerical simulations show that the higher amplitude circular component is amplified up to the point where it reaches a steady-state solution with an amplitude given by Eq.(15.3). Whereas the pulse component is the other circular polarization also evolves towards a steady-state pulse shape with an amplitude such that the initial amplitude ratio (or ellipticity of the pulse polarization state) is maintained unchanged. Moreover we observe that, at steady-state, the ratio of among the asymptotic populations in the two circular components is equal to the square of the initial amplitudes ratio. For example, whenever $e_{20}/e_{10} = 3$, then $n_{11} \rightarrow 0.05$ and $n_{22} \rightarrow 0.45$ at $\tau \rightarrow +\infty$. In other words, initial weak pulses of the form (18) are amplified and ultimately evolve into a steady-state vector π-pulse, that stably propagates in the host fiber. In these simulations, for simplicity we neglected the presence of linear birefringence and the GVD of the host medium. The process of attraction of a weak input pulse into a stable vector π-soliton is illustrated in figure 4: Here the amplitudes of the initial pulse $e_{1,2}(\zeta = 0, \tau)$ are set as one half (one quarter) of the steady-state value, respectively. It is interesting to observe that the coherent amplification process leads to a steady state vector π-pulse which maintains the input ratio between the two circular polarization components of the initial small pulse. In other words, in the absence of linear birefringence, the coherent amplification process does not change the input polarization state of a weak pulse (see Eqs.(8.4-8.5).

Finally, let us briefly consider the effect of fiber birefringence in the creation from a small pulse of vector π-pulses in a fiber with inverted resonant atoms. As it can be seen from figure 5, the birefringence does not prevent the formation of a stable vector π-pulse whose polarization state periodically rotates on the Poincaré sphere about the linear birefringence axes. Note in figure 5 that the polarization rotation is accompanied by a spatially periodic modulation of the atomic inversion of the atoms that interact with either the left or the right-handed circular polarization components of the light pulse.

## Conclusions

We investigated the amplification and propagation of the short electromagnetic solitary waves in a linearly birefringent and dispersive optical fiber containing resonant atoms with an inverted population of the energy levels. The resonant medium model involved two-level atoms, with an upper state that is degenerate on the projection of the angular moment. In the case of an isotropic fiber and when GVD can be neglected, we obtained an analytic vector generalization of the scalar $\pi$-pulse of light which results from the balance between linear absorption of the host fiber and nonlinear amplification by the active atoms. The polarization state of the solitary wave reflects the specific properties of the resonant atomic system, as well as the polarization of the pump wave. A peculiar property of the linearly polarized vector $\pi$ solution is that after the interaction with the optical pulse the atomic inversion does not vanish, that is the pulse does not leave the atoms in their ground state. We further investigated and confirmed by means of simulations the robustness of the linearly polarized vector $\pi$-pulse in the presence of both birefringence and dispersion of the host fiber. Moreover, we numerically observed that a weak incident pulse of arbitrary polarization is amplified until it is transformed into a stable vector $\pi$ pulse without a change of its initial polarization state.

## Acknowledgements

One of the authors (E.V.K.) is grateful to the Laboratoire de Physique de l'Université de Bourgogne (L.P.U.B) Unité Mixte de Recherche du CNRS 5027, Faculté des Sciences, Dijon for hospitality and support and to the Conseil Régional de Bourgogne for funding of her postdoc fellowship.

## Figure Captions

Fig.1. Vector steady-state solution for both circular polarization components of the electric field and density matrix elements for the case of an isotropic fiber. The length of resonant absorption is equal to $l_a$ =0.5.

Fig.2. Evolution of the circular polarization components of a an initial vector π-pulse at 90 degrees to the axes of a birefringent fiber with $l_c$ =2.

Fig.3. Same as in Figure 2, with $l_c$ =0.5.

Fig.4. Vector steady state pulse creation from a small initial pulse.

Fig.5. Creation and polarization rotation of a vector dissipative soliton in a birefringent fiber with $l_c$ =1.

Fig.6. Robustness of vector π-pulse in the presence of weak host dispersion ($l_d$ =5).

Fig.7. Distortion of vector π-pulse in the presence of strong host dispersion ($l_d$ =1).

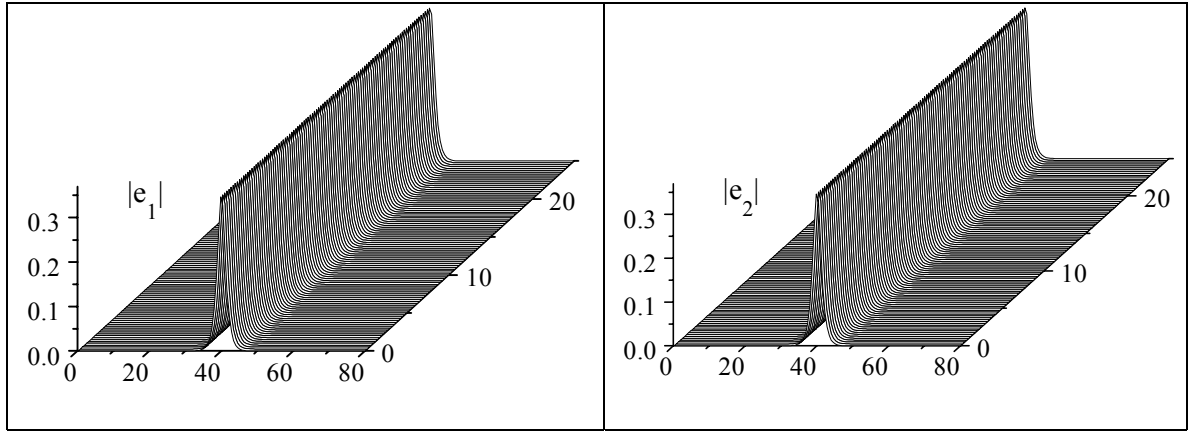

**Figure 1**

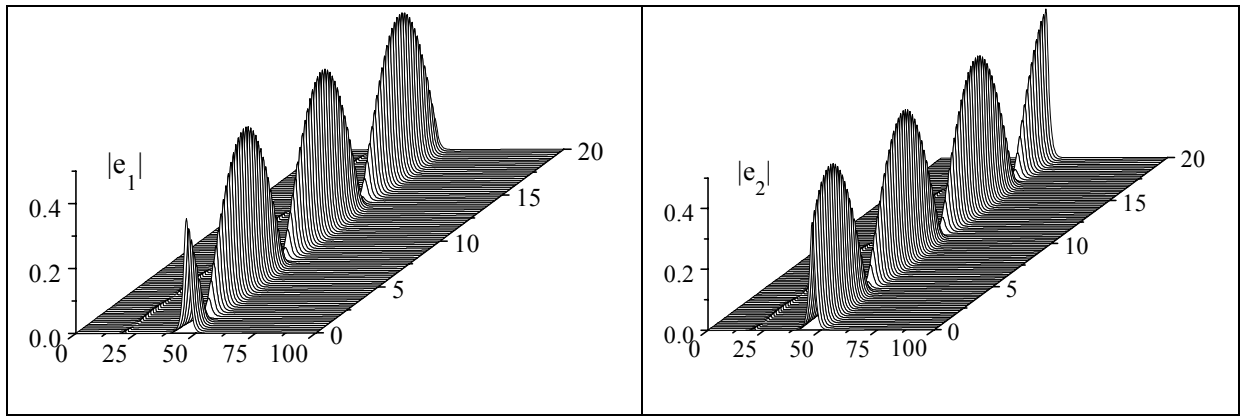

**Figure 2**

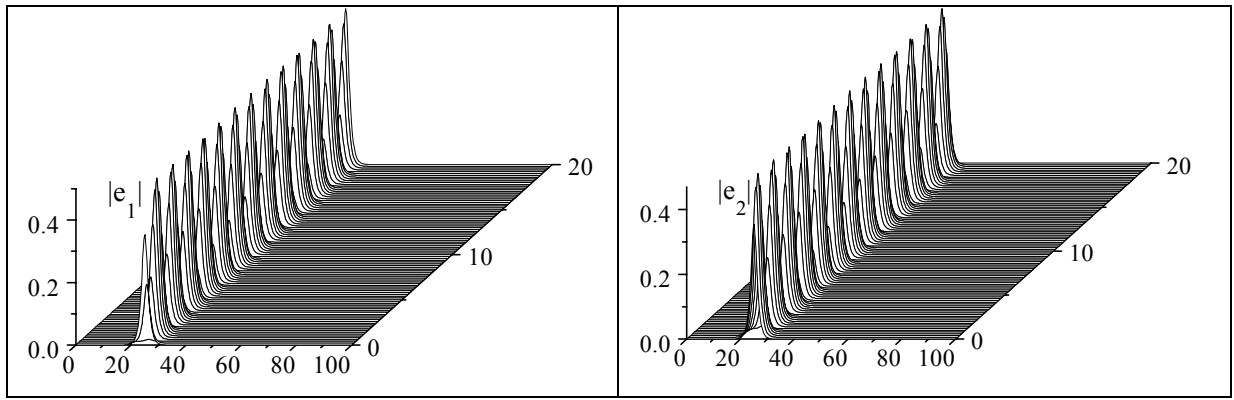

**Figure 3**

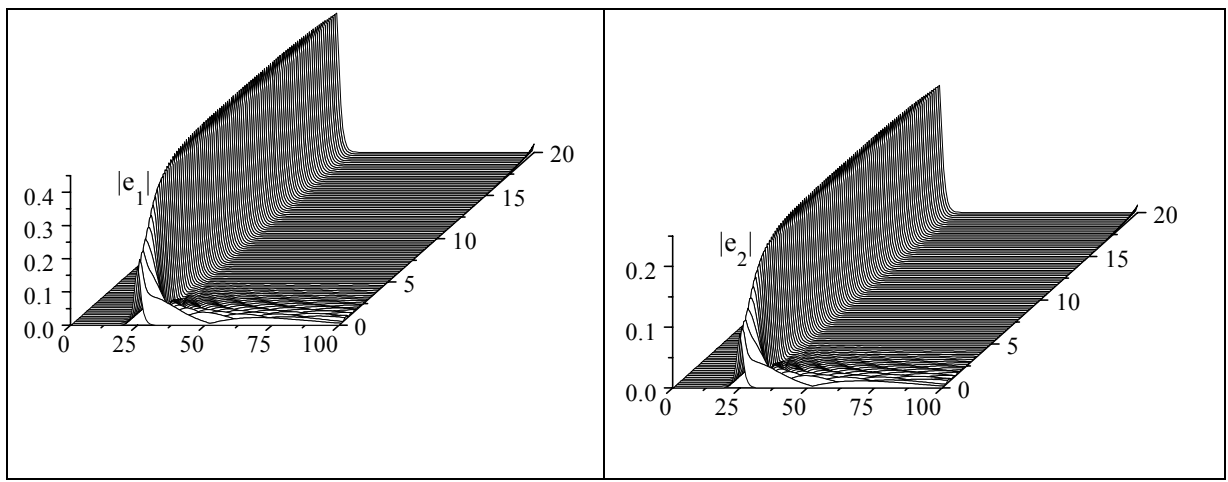

**Figure 4**

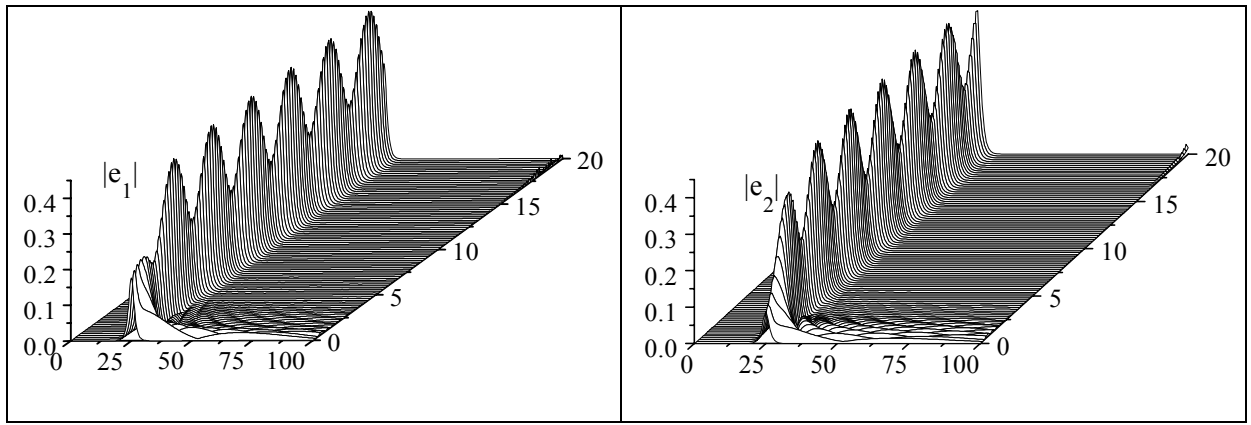

**Figure 5**

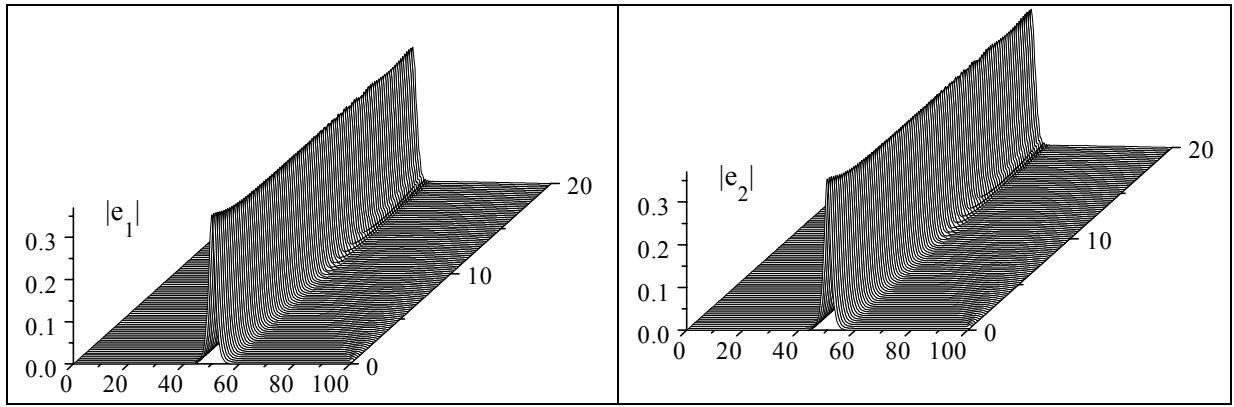

**Figure 6**

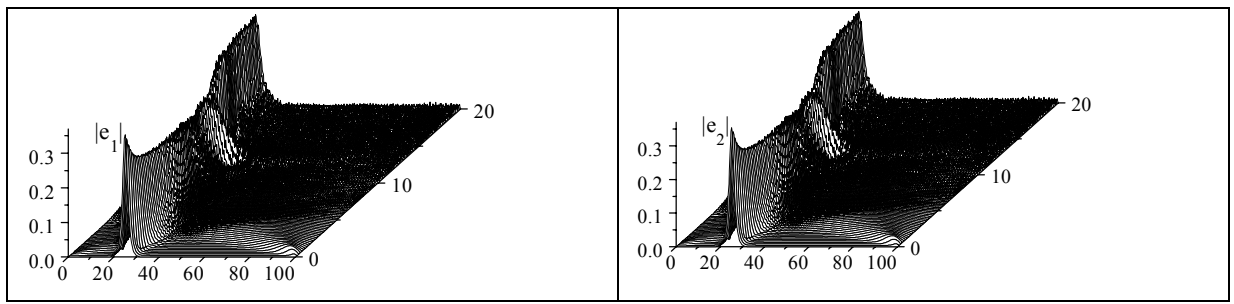

**Figure 7**